# Spectra and waiting-time densities in firing resonant and nonresonant neurons

T. Verechtchaguina, L. Schimansky-Geier, and I. M. Sokolov
*Institute for Physics, Humboldt University of Berlin, Newton Strasse 15, D-12489 Berlin, Germany*

The response of a neural cell to an external stimulus can follow one of two patterns: nonresonant neurons monotonically relax to the resting state after excitation while resonant ones show subthreshold oscillations. We investigate how these subthreshold properties of neurons affect their suprathreshold response. Conversely we ask the following: Can we distinguish between both types of neuronal dynamics using suprathreshold spike trains? The dynamics of neurons is given by stochastic FitzHugh-Nagumo and Morris-Lecar models either having a focus or a node as the stable fixed point. We determine numerically the spectral power density as well as the interspike interval density in response to random (noiselike) signals. We show that the information about the type of dynamics obtained from power spectra is of limited validity. In contrast, the interspike interval density provides a very sensitive instrument for the diagnostics of whether the dynamics has resonant or nonresonant properties. For the latter value, we formulate a fit formula and use it to reconstruct theoretically the spectral power density, which coincides with the numerically obtained spectra. We underline that the renewal theory is applicable to analysis of suprathreshold responses even of resonant neurons.

## I. INTRODUCTION

The response of a neural cell to an external stimulus can follow one of two patterns: Cells of the first type (nonresonant neurons) monotonously relax to the resting state after excitation; neurons of the second type (resonant neurons) show subthreshold oscillations at a well-prescribed frequency.

Evidence of subthreshold oscillations is experimentally shown in mesencephalic trigeminal neurons [1], dorsal root ganglion neurons [2], neocortical neurons [3,4], thalamic neurons [5], and others. To detect subthreshold oscillations in experiment, a potential on the weakly depolarized neuron membrane is usually recorded in the subthreshold regime when the neuron does not fire. A peak in the power spectrum of the recorded signal and the periodicity of its autocorrelation function serve as indicators of subthreshold oscillations [2,6,7].

The presence of a preferred frequency in neuronal dynamics may lead to selective subthreshold and suprathreshold responses with respect to the input of periodic pulse sequences or of noisy signals. A resonant peak appears in the impedance for subthreshold response to signals sweeping through many frequencies over time (the so-called ZAP input) [1,8].

The importance of studying suprathreshold response in firing neurons is connected with the fact that frequency preference contributes to synchronization of electrically coupled neurons [9,10], which is assumed to be important for neuronal processing [11]. Subthreshold oscillations trigger spikes following with the preferred frequency in suprathreshold response [12,10,13], so that the bursting effect and the firing-rate resonance are present. The relation between subthreshold oscillations and firing-rate resonance is investigated theoretically in [14–17] in the generalized integrate-and-fire model [15,18]. In experimental studies, spectral analysis is normally used for quantitative characterization of the oscillatory activity under various experimental conditions [9,7,13].

However, resonant and nonresonant neurons in response to external signals generate spike trains, which look rather similar if observed on a shorter time scale. Looking at longer scales and especially considering the power spectra of these outputs reveals some differences. However, these are not so striking in the *form* of spectra; rather, they are remarkable in a different response to *changes of parameters* of the inputting signal. As will be seen, the assessment of dynamics based on such spectra is of limited validity.

As we proceed to show, the density of the interspike interval distribution provides a much more sensitive instrument for this assessment since the differences between both regimes are clearly seen without additional data processing. Moreover, it is known that this density contains practically all relevant information, so that the power spectrum can be restored from it and from the spectrum of a single spike.

In order to complete our task, we need a theoretical model of a neuron which can readily reproduce both regimes (resonant and nonresonant) and is simple enough to allow for massive numerical simulations necessary for adequate statistical analysis. For such a model we have chosen a FitzHugh-Nagumo (FN) system [19,20] and consider its response to a signal which is modeled by a white noise. This kind of noisy input allows for a mathematically sound formulation of the problem and for use of approved tools of stochastic modeling; from the physiological point of view, it corresponds to a signal from a complex environment changing on time scales which are shorter than the time of the neuron's response. This is often the case for real biological systems [21]. Another model we used was the Morris-Lecar model [22].

## II. SUBTHRESHOLD DYNAMICS OF THE FITZHUGH-NAGUMO MODEL

The deterministic FitzHugh-Nagumo (FN) model [19,20] is a formal model describing the neuron dynamics in terms of only two relevant variables: the voltage variable $x(t)$ and the recovery variable $y(t)$, which represent the effective membrane conductivity,

$$\epsilon \dot{x} = x - x^3 - y, \quad \dot{y} = \gamma x - y + b. \quad (1)$$

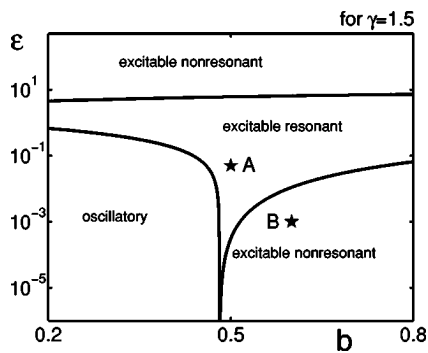

FIG. 1. Stability diagram of the FN model in parameter space ($\epsilon, b$) for constant $\gamma = 1.5$; see text for details. Points $A$ and $B$ denote parameter sets used in this paper to model resonant (point $A$: $\epsilon = 0.05$, $\gamma = 1.5$, $b = 0.5$) and nonresonant (point $B$: $\epsilon = 0.001$, $\gamma = 1.5$, $b = 0.6$) neurons. Note that in all the figures, plotted quantities are dimensionless.

The parameter $\epsilon > 0$ (representing the membrane capacity) is an important parameter of the model. It is a small positive constant ($\epsilon \ll 1, \epsilon \ll 1/\gamma$), which guarantees the time-scale separation. The time evolution of $x(t)$ reproduces all qualitative features of spike generation on neuron membrane. All parameters and variables of the FN model are dimensionless. Depending on the parameters $\gamma$ and $b$, the system can show three different regimes of dynamical behavior. These are excitable, oscillatory, and bistable regimes [23]. In what follows, we consider the system only in the excitable regime, where the system possesses one stable fixed point in the phase space.

In what follows, we set $\gamma = 1.5$, and carry out the stability analysis of the system (1) in parameter space ($\epsilon, b$). Results are presented in Fig. 1. In the excitable nonresonant regime, the fixed point is a stable node with the Lyapunov exponents being real and negative (Re$\lambda < 0$, Im$\lambda = 0$). For parameter values from the region corresponding to the resonant excitable regime, the fixed point is a stable focus; the Lyapunov exponents are two complex conjugates (Re$\lambda < 0$, Im$\lambda \neq 0$). Points $A$ and $B$ denote two parameter sets, which we use in this paper to model different types of neuronal behavior: point $A(\epsilon = 0.05, \gamma = 1.5, b = 0.5)$ corresponds to resonant and point $B(\epsilon = 0.001, \gamma = 1.5, b = 0.6)$ to nonresonant neurons. Note that the resonant regime can be obtained for finite values of $\epsilon$ only; the limit $\epsilon \to 0$ corresponds to the nonresonant case.

Subthreshold dynamics of the FN model for these two parameter sets is illustrated in Figs. 2(a) and 2(b). Subthreshold oscillations in the resonant case and monotonic behavior in the nonresonant one in the time evolution of the voltage variable are clearly seen. In the resonant case, there are two intrinsic characteristic times in the system besides the refractory time [Fig. 2(a)]. These are the time between the beginning of a spike and the first maximum following subthresh-old oscillation ($T_1 = 2\pi/\omega_1$, $T_1 \sim 2.5$, $\omega_1 \sim 2.4$; all quantities are dimensionless) and the period of subthreshold oscillation ($T_2 = 2\pi/\omega_2$, $T_2 \sim 1.12$, $\omega_2 \sim 5.6$).

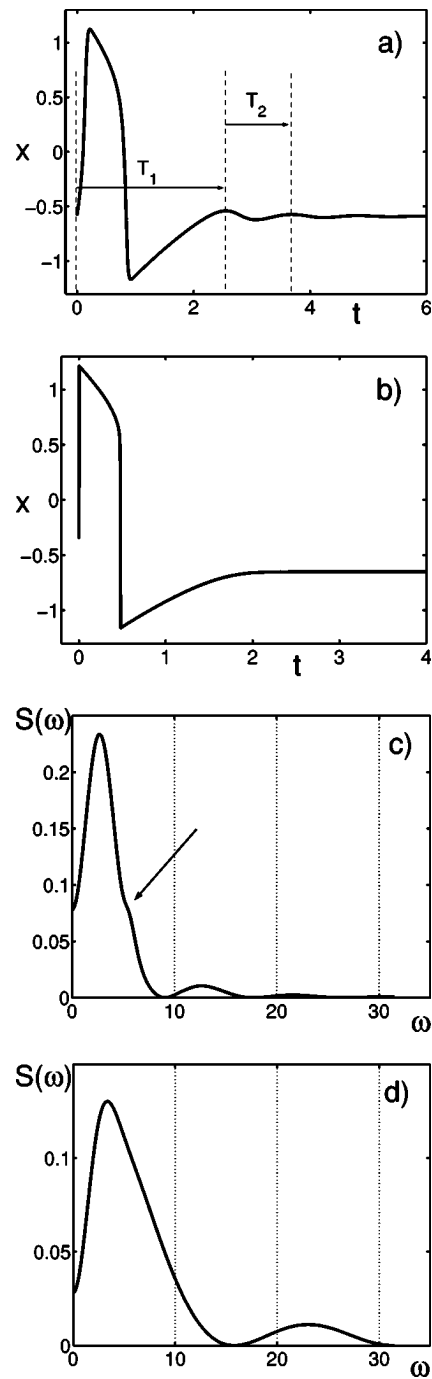

FIG. 2. (a),(b) Spike shape in the FN model [the voltage variable as a function of time $x(t)$; note the difference in the time scales]. (c),(d) Power spectrum of a single spike in the FN model without noise $D = 0.0$; the spectrum is normalized so that the integral over the spectral density is unity. (a),(c) The resonant regime ($\epsilon = 0.05, \gamma = 1.5, b = 0.5$), subthreshold oscillations. (b),(d) The nonresonant regime ($\epsilon = 0.001, \gamma = 1.5, b = 0.6$).

After a spike is generated, the following subthreshold oscillation modulates the distance to the excitation threshold, i.e., the probability to generate the next spike. This probabil-ity is close to zero during the refractory period. It achieves its maximum when the membrane potential goes through the first maximum of subthreshold oscillation, i.e., $T_1$ is the most probable interspike interval (ISI). If the second spike is generated in this time, one speaks about a spike burst; spikes follow with the frequency $\omega_1$. If there is no second spike on $T_1$, there is a higher probability that it will be generated at the second maximum of subthreshold oscillation, so that the ISI will be $T_3 = T_1 + T_2 \sim 3.6$ and the corresponding frequency is $\omega_3 = 2\pi/T_3 \sim 1.7$, and so on. Hence these intrinsic characteristic times are expected to be reflected in the suprathreshold responses of the resonant neurons to external signals.

Power spectra of a single spike in the resonant and nonresonant regimes are presented in Figs. 2(c) and 2(d). Spectra are obtained numerically and normalized so that the total signal power equals unity. In the resonant case, one can see a tiny shoulder in the spectrum on the frequency close to $\omega_2$, caused by the subthreshold oscillation. However, the magnitude of the shoulder is so small that in stochastic spectra of neuron responses to a noiselike input it will not be distinguishable in the noise background.

In an *in vivo* environment, a neuron receives a continuous barrage of inputs arriving at a large number of synapses [17]. This situation can be modeled by a white Gaussian noise $\xi(t)$ of intensity $D$, $\langle \xi(t)\xi(t')\rangle = \delta(t-t')$, leading to a FN model with a stochastic input. The common representation of the stochastic FN model reads [21]

$$\epsilon \dot{x} = x - x^3 - y, \qquad (2)$$

$$\dot{y} = \gamma x - y + b + \sqrt{2D}\xi(t).$$

The white noise term in the second equation of the system of Eqs. (2) can be turned to a colored noise term in the first equation by a change of variables [23]. The difference between these cases is discussed in detail in [23]. Including the white noise term in the second equation in the system of Eqs. (2) is the most common and mathematically consequent way to add stochastic input to the FN model (see [21,24,25] and the discussions therein). Due to the noise, the FN model generates spike sequences; see Fig. 3. It is complicated to distinguish between two regimes just observing such signals during short time intervals. This makes it necessary to introduce reliable statistical instruments which distinguish clearly between the two situations.

### III. POWER SPECTRAL DENSITY

It is customary to use power spectra to distinguish between resonant and nonresonant phenomena. The power spectrum $S_x(\omega)$ of the voltage variable $x(t)$ is defined as

$$S_x(\omega) = x(\omega)x^*(\omega), \qquad (3)$$

with $x(\omega)$ being the Fourier transform of the corresponding variable,

$$x(\omega) = \int_{-\infty}^{+\infty} x(t)e^{i\omega t}dt.$$

Spectra of the stochastic FN model were calculated analytically in [23,25]. However, the approximation which made the analytical approach possible corresponds to the limit $\epsilon \to 0$. Hence only nonresonant neurons could be adequately described.

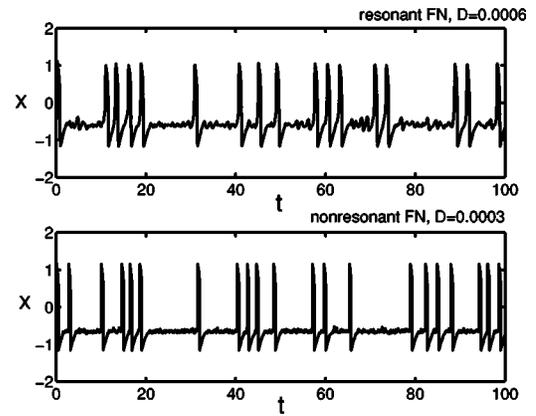

FIG. 3. Pulse trains obtained from the FN model. Top: resonant regime, $D=0.0006$, bursting effect; bottom: nonresonant regime, $D=0.0003$. The values of noise intensities are chosen in such a way that the mean density of spikes is approximately the same.

Let us first turn to the numerically obtained spectra of the FN model. To get them, we sampled a signal from stochastic FN output at $2^{15}$ points with a sample interval 0.1 and calculated power spectral density using the fast Fourier transform [26]. The spectra were averaged over 120 different realizations. Results are presented in Fig. 4 for the resonant and nonresonant regimes for different noise intensities. The normalization is the same as in Fig. 2.

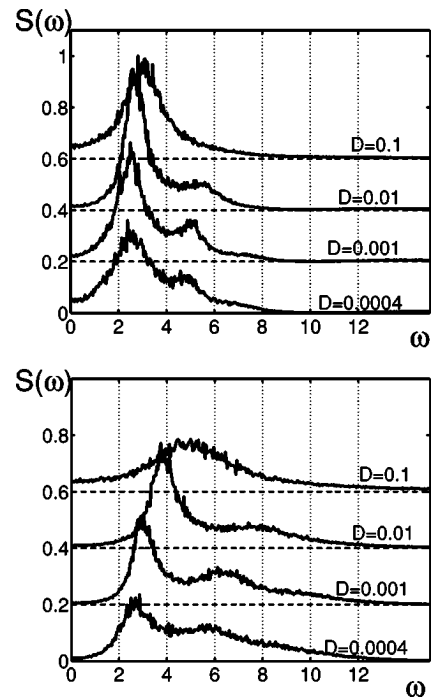

FIG. 4. Spectrum $S(\omega)$ of the stochastic FN neuron in the resonant (top) and nonresonant (bottom) regimes for $D=0.0004, D=0.001, D=0.01, D=0.1$; see text for details. The spectra for different noise intensities are vertically shifted for the sake of clarity.

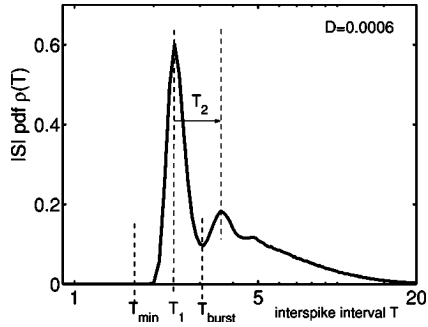

FIG. 5. Interspike interval PDF obtained from simulations for the stochastic FN model in the resonant regime for $D=0.0006$ as obtained from $10^6$ interspike intervals. Note the logarithmic scales, making evident the oscillatory behavior of resonant PDF.

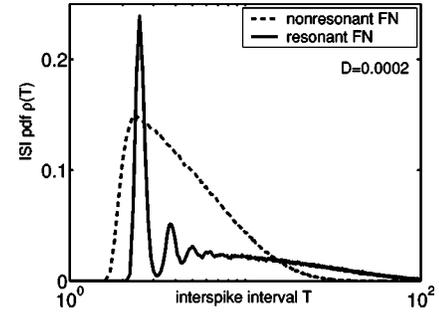

FIG. 6. Interspike interval PDF obtained from simulations of the stochastic FN model in resonant and nonresonant regimes for the same noise intensity $D=0.0002$. Note the logarithmic scales.

The notable features of the spectra are the existence of a well-pronounced main peak on the mean firing frequency and a lower peak at higher frequency which disappears at high levels of noise. With increasing noise, the main peak moves to higher frequencies due to a decrease of the mean interspike interval. However, this shift is much less pronounced in the resonant regime: the frequency corresponding to this peak stays close to $\omega_1$.

Besides this fact and some quantitative differences, resonant and nonresonant spectra are very similar. The second peak moves to higher frequencies as well so that its frequency remains approximately double that of the first peak's frequency, while the frequency of subthreshold oscillation remains constant. This fact and the fact that the second peak is present in both resonant and nonresonant spectra are evidence of its nature as the second harmonic.

## IV. WAITING-TIME DENSITY

The probability density function (PDF) of interspike intervals is obtained from a neuronal output if reduced to a $\delta$-spike sequence. The whole dynamics between spikes is neglected, thus the waiting-time density is a very simplified function at first glance. Nevertheless, as will be seen, this waiting-time density contains almost all the relevant information. In particular, the spectra of the output can be reconstructed from the waiting-time density by renewal theory and from the spectrum of a single spike (the form factor, Fig. 2).

The waiting-time density for the stochastic FN model is obtained numerically. We have simulated stochastic FN equations (2) in the resonant ($\epsilon=0.05, \gamma=1.5, b=0.5$) and nonresonant ($\epsilon=0.001, \gamma=1.5, b=0.6$) regimes for different noise intensities $D$. The spikes were defined as zero-level crossings from negative to positive values of the $x$ variable. We collected $10^6$ interspike intervals $T$ to obtain waiting-time (i.e., interspike interval) PDF $\rho(T)$.

Results of simulations are presented in Fig. 5 for resonant FN neuron for $D=0.0006$ and in Fig. 6 for resonant and nonresonant FN neurons ($D=0.0002$). The ISI probability density functions differ qualitatively in these two regimes: $\rho(T)$ for a nonresonant neuron possesses only one maximum, while the PDF in the resonant case shows oscillations. The first very sharp maximum in the latter case occurs near $T_1$ and corresponds to the ISI within a burst. Further maxima follow with the period of subthreshold oscillations $T_2$ (in Fig. 5 the characteristic times $T_1$ and $T_2$ are the same as in Fig. 2). The tail of the resonant ISI PDF is flat and long, thus the bursts are separated by large, almost uniformly distributed intervals. A very natural definition of a burst is then that it is a sequence of spikes separated by intervals smaller than the time $T_{\text{burst}}$ up to the first *minimum* of the ISI PDF (Fig. 5).

To understand the oscillatory behavior of the resonant ISI PDF, we can interpret it as follows: Assume the spike generation process to be the renewal one and the first spike is generated on the time $T=0$. Then $\rho(T)$ is the time-dependent probability that the next spike is generated on time $T$. This probability is modulated in the resonant case by the subthreshold oscillation.

As mentioned, nonresonant waiting-time density lacks both the oscillatory behavior and the long tail. Therefore, the neuronal output is more homogeneous in this case. Note, finally, that there is a certain minimal value $T_{\min}$, such that for all $T<T_{\min}$ the interspike interval PDF vanishes. This value $T_{\min}$ is the relative refractory time, which decreases with increasing noise intensity.

## V. POWER SPECTRAL DENSITY OBTAINED FROM WAITING-TIME DENSITY

Let us see how the spectrum of the FN neuron can be obtained from numerical data for waiting-time density. The idea to use the numerically obtained PDF in theoretical considerations was already proposed in [24]. In our case, a combination of the numerical and analytical methods allows us to avoid a lot of simplifying assumptions and to obtain the power spectrum even in the resonant case. The idea is as follows.

(i) Instead of examining the complete dynamics of the stochastic FN neuron, we restrict our consideration to "events" (spikes) only. The output $x(t)$ of a FN neuron is replaced with a point process

$$\sigma(t) = \Sigma \delta(t-t_i). \tag{4}$$

The spikes occur at times $t_i$ in the FN output $x(t)$.

(ii) Assuming the spike generation process is a *renewal process*, we apply a Stratonovich formula for the renewal

TABLE I. The values of fit parameters.

| $D$ | $T_1$ | $T_{\min}$ | $u_0$ | $u_1$ | $u_2$ | $u_3$ | $u_4$ |
|---|---|---|---|---|---|---|---|
| | | | Resonant | | | | |
| 0.0006 | 2.1 | 2.1 | 0.582 | 0.564 | 4.44 | 0.199 | 4.517 |
| 0.002 | 2.1 | 1.9 | 0.479 | 0.457 | 5.383 | 0.579 | 1.488 |
| 0.02 | 2.1 | 1.54 | 0.222 | 0.298 | 2.736 | 0.758 | 0.835 |
| 0.09 | 2.0 | 1.07 | 0.237 | 0.435 | 2.061 | 0.632 | 0.847 |
| | | | Nonresonant | | | | |
| 0.0002 | 2.2 | 1.76 | 0.132 | 0.329 | 0.717 | 0.148 | 6.463 |
| 0.002 | 2.0 | 1.3 | 0.098 | 0.179 | 3.473 | 0.857 | 0.522 |
| 0.03 | 1.5 | 0.8 | 0.276 | 0.244 | 3.045 | 0.923 | 0.461 |
| 0.1 | 1.4 | 0.5 | 0.193 | 0.268 | 2.44 | 0.728 | 0.489 |

sequence of $\delta$ spikes, which connects the power spectral density and the characteristic function $\rho(\omega)$ of the interspike interval $T$ [27],

$$S_\sigma(\omega) = \frac{1}{\langle T \rangle} \frac{1-|\rho(\omega)|^2}{|1-\rho(\omega)|^2} + \frac{2\pi}{\langle T \rangle^2} \delta(\omega). \quad (5)$$

$\langle T \rangle$ is the mean interspike interval,

$$\langle T \rangle = \lim_{N \to \infty} \frac{1}{N} \sum_{i=1}^{N} T_i, \quad (6)$$

and the characteristic function $\rho(\omega)$ is the Fourier transform of the interspike interval PDF $\rho(T)$,

$$\rho(\omega) = \int_0^\infty \rho(T) e^{i\omega T} dT. \quad (7)$$

Analogous calculations of power spectra based on the Stratonovich formula were made for the leaky integrate-and-fire model [23,28] and for the nonresonant FitzHugh-Nagumo model [18].

In contrast with these approaches, in the present work we obtain the interspike interval PDF from simulations (Figs. 5 and 6) and fit it to a function, which is then substituted into Eqs. (7) and (5)

In particular, we used to fit $\rho(T)$,

$$\rho(T) = \Theta(T-T_{\min})[u_0 e^{-(T-T_1)/u_1} \sin[u_2(T-T_1)] + u_3 e^{-(T-T_1)/u_4}]. \quad (8)$$

Here $\Theta(T-T_{\min})$ is the Heaviside function: there are no interspike intervals smaller than $T_{\min}$. $T_1$ is the time between the beginning of a spike and the first maximum of the following subthreshold oscillation; it is the most probable interspike interval. These two parameters can be found immediately from simulation data. $u_0, u_1, u_2, u_3, u_4$ are proper fit parameters. $u_2$ denotes the frequency of subthreshold oscillations and $u_4$ is a characteristic time of the barrier crossing mechanism. $u_0$ and $u_3$ are weight parameters. Actually, $u_0, u_1, u_2, u_3, u_4$ are not independent because of the normalization condition $\int_0^\infty \rho(T) dT = 1$.

The values of $u_1, u_2, u_3, u_4$ must be found for every noise intensity. To fit numerical data to function Eq. (8), we used the Levenberg-Marquardt method [26]. Obtained values of the fit parameters are presented in Table I for some noise intensities. The Kolmogorov-Smirnov test [26] (the most sensible one for cumulative distribution functions) confirms a good fit quality with a confidence level better than 0.99. The Fourier transform of function Eq. (8) can be derived exactly as a function of fit parameters.

The applicability of this method is restricted by the amount of available data. For too small data sets, the form of the ISI PDF depends strongly on the interval that was used to collect the PDF. This makes an appropriate fit impossible. The estimated minimal number of ISI is about 1000. Provided with a limited amount of experimental data, one can still apply this method: One has to consider the cumulative distribution function, which is independent of the sampling interval, and find a fit for it. After that, the ISI PDF and spectra can be derived analytically.

The final results for the spectra $S(\omega)$ are presented in Fig. 7 for noise intensities listed in Table I. Let us pay attention to the asymptotic behavior of these spectra in the high- and low-frequency domains. For high frequencies, the spectrum $S(\omega)$ of the $\delta$ spikes saturates at a level related to the stationary firing rate $r_0$,

$$\lim_{\omega \to \infty} S(\omega) = r_0, \quad (9)$$

whereby $r_0 = 1/\langle T \rangle$ [23,27]. This effect is connected with the fact that the $\delta$-spike train possesses an infinite variance: $\lim_{\tau \to 0} \langle \sigma(t) \sigma(t+\tau) \rangle = r_0 \delta(\tau)$ [23]. For low frequencies it can be shown [23,28] that

$$\lim_{\omega \to 0} S(\omega) = R^2 r_0, \quad (10)$$

with the coefficient of variation (CV) $R = \sqrt{\langle \Delta T^2 \rangle}/\langle T \rangle$.

If the CV is below 1, then $S(\omega)$ in the low-frequency limit is smaller than $S(\omega)$ in the high-frequency limit. This low-frequency dip is related to the refractory period. For small noise intensities, mean excitation time is very large and refractory time becomes irrelevant, thus the spectrum approaches the flat Poisson spectrum ($R \to 1$) with a weak dip. For intermediate noise, when refractory and escape times are of the same order, refractoriness determines firing rate, and neuronal output becomes most regular. This is the effect of coherence resonance [29]; the CV is small in this case and the low-frequency dip in the spectrum is strong.

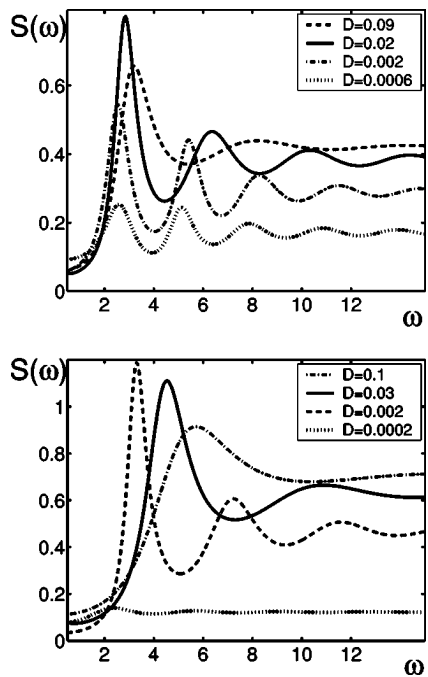

FIG. 7. Spectrum $S(\omega)$ of a $\delta$ output of a stochastic FN neuron for different noise intensities, in the resonant regime ($D=0.0006$, 0.002, 0.002, 0.02, 0.09) (top), and in the nonresonant regime ($D=0.0002$, 0.002, 0.03, 0.1) (bottom).

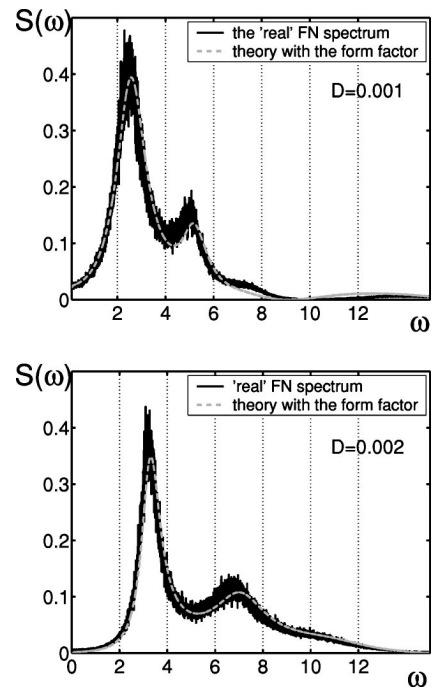

FIG. 8. Power spectrum of the stochastic FN neuron calculated numerically (black line) and power spectrum calculated theoretically and multiplied with the form factor (gray dashed line); top: resonant regime, $D=0.001$; bottom: nonresonant regime, $D=0.002$. Spectra are normalized on the total signal power.

Moreover, Fig. 7 shows that in the resonant case the dip is weaker than in the nonresonant one, especially for a weak noise. In the nonresonant regime, spike trains are more homogeneous, while in the resonant regime, spike bursts separated by long intervals are typical, which results in higher values of CV.

A visual inspection of the spectra theoretically calculated using numerical data for the waiting-time density (Fig. 7) shows that the height of the first peak, the frequency $\omega_{max}$ where the power spectrum attains its maximum, and the peak's shift to higher frequencies with increasing noise intensity are analogous to those of the spectra estimated numerically (Fig. 4). In the resonant regime, $\omega_{max}$ also remains close to the frequency $\omega_1$ with increasing noise intensity.

Nevertheless, so far the spectra look very different from the numerically determined ones. They contain considerably more higher harmonics, and they do not vanish in the limit of high frequencies. This can be explained by the fact that $\delta$ spike sequences instead of the whole FN output were used so far. This output generated by the FN model is a convolution of the single-spike form with the $\delta$ sequence indicating the spikes' positions (here the forms of the pulses are assumed to be identical). Hence the spectrum of the $\delta$ spike multiplied with the form factor (which is the spectrum of a single spike) should coincide with the spectrum calculated numerically. However, by this reasoning we neglect all interspike dynamics, which contains the subthreshold oscillation on the well-prescribed frequency in the resonant case. If this frequency were directly reflected in the spectrum, for example as a peak on it, then the spectrum calculated as a spectrum of a point process would lack this peak on the frequency of the subthreshold oscillation. Thus the coincidence of spectra obtained theoretically and numerically is possible only if the frequency of subthreshold oscillation is not directly presented in the power spectrum.

We have calculated numerically power spectra of single spikes in autonomous (without noise, $D=0.0$) FN system in both regimes (single-pulse form factors). The results of the multiplication of theoretically calculated spectra with the form factors are presented in Fig. 8 (gray dashed line). In the same figure, numerically estimated spectra are depicted by a black line. One can note a good agreement between numerical and theoretical results.

This agreement confirms our assumption that spike generation is a renewal process. The assumption also holds in the case of resonant neurons, which justifies the application of tools from the theory of point process in order to reconstruct the power spectrum. We again find that the frequency of subthreshold oscillation is not directly reflected in the power spectrum density.

The small difference between the numerical and the semi-theoretical results might be caused by the change of the spike form if external noise is applied. This difference is evident for a strong noise ($D=0.1$) and indiscernible for small noise intensities. All higher harmonics observable in spectra calculated from the waiting-time density (Fig. 7) disappear after multiplication by the form factor.

We repeated all simulations and calculations for the Morris-Lecar model,

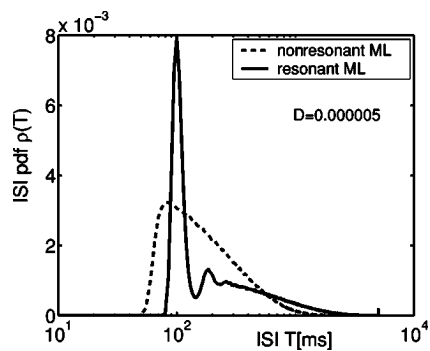

FIG. 9. Interspike interval PDF obtained from simulations of the stochastic Morris-Lecar model ($V_1=-1.2, V_2=18, V_3=2, V_4=30, \bar{g}_{Ca}=4.4, \bar{g}_K=8.0, \bar{g}_L=2$, $V_K=-84$, $V_L=-60$, $V_{Ca}=120, \phi=0.04$) in resonant ($C=20, I=82$) and nonresonant ($C=2, I=65$) regimes for the same noise intensity $D=0.000\,005$. Note the logarithmic scales.

$$C\frac{dV}{dt} = -\bar{g}_{Ca}m_\infty(V)(V-V_{Ca}) - \bar{g}_K w(V-V_K) - \bar{g}_L(V-V_L) + I, \quad (11)$$

$$\frac{dw}{dt} = \phi\frac{[w_\infty(V)-w]}{\tau_w(V)},$$

where $m_\infty(V)=0.5*[1+\tanh(V-V_1)/V_2]$, $w_\infty(V)=0.5*[1+\tanh(V-V_3)/V_4]$, and $\tau_\infty(V)1/\cosh(V-V_3)/(2*V_4)$. Parameter values used to simulate the resonant and nonresonant regimes are given in the legend to Fig. 9.

In Fig. 9, the ISI PDF for a stochastic Morris-Lecar model is presented in the resonant and nonresonant regimes for $D=0.000\,005$. The results are quite analogous to those presented by the FN model, what argues for the generality of the results.

## VI. SUMMARY

We discussed the spectral properties and the waiting-time densities of the FitzHugh-Nagumo model subjected to white Gaussian noise. We show that there is no pronounced qualitative difference between power spectra of resonant and of nonresonant regimes. The difference between spectra of resonant and nonresonant neurons is mostly quantitative. A small distinction is found in their response to a possible change of the external driving. But, although spectral characteristics are very habitual and widely used, they are not so informative, as we have seen from simulations of a surrogate mathematical model of a neuron.

In contrast, the form of the interspike interval probability density function differs strongly for these two cases. Thus the ISI PDF in the resonant case shows an oscillatory behavior, which mirrors subthreshold oscillations, while the ISI PDF in the nonresonant case possesses only one maximum. Moreover, the structure of resonant ISI PDF assumes a bursting effect, actually observed in resonant neurons, and gives an accurate definition of bursts. For the considered FitzHugh-Nagumo model, we formulated a fit formula for the ISI PDF.

Using the Stratonovich formula for spectra of point processes, we reconstructed from these waiting-time densities the power spectra of the output after convolution with the form factor of a single neuron. We found good agreement of these spectra with the numerically estimated ones. Therefore, the information contained in power spectra can be extracted from the waiting-time density and the spectrum of a single spike, and the assumption of a renewal process works as well for resonant neurons.

As a result, the interspike interval PDF contains almost all relevant information even in the resonant case. We may conclude that the waiting-time density rather than power spectra has to be preferred in experiments of the neuronal response.

## ACKNOWLEDGMENTS

We acknowledge financial support from the DFG though Graduierten Kolleg 268 and Sfb 555. I.M.S. acknowledges partial support from the Fonds der Chemischen Industrie.